# The 'Problematic Paper Screener' automatically selects suspect publications for post-publication (re)assessment

Guillaume Cabanac · Cyril Labbé · Alexander Magazinov

**Objective:**
Post publication assessment remains necessary to check erroneous or fraudulent scientific publications. We present an online platform, the 'Problematic Paper Screener' (PPS) that leverages both automatic machine detection and human assessment to identify and flag already published problematic articles. We provide a new effective tool to curate the scientific literature.

**Method:**
PPS combs the scientific literature for a variety of research integrity issues. Malpractices are automatically identified thanks to specific 'fingerprint-queries' submitted to the academic search engine Dimensions.ai. Peer judgement can then confirm (or refute) the status of suspect papers. The public online interface allows public users to propose new 'fingerprints'.

**Results:**
As of today (Oct. 2021) the PPS (https://www.irit.fr/~Guillaume.Cabanac/problematic-paper-screener) lists papers containing meaningless computer-generated texts (N=264 SCIgen, 11 Mathgen, 4 SBIR). For this kind of malpractice, 'fingerprint-queries' are specific sequences of words extracted from the probabilistic context free grammars like 'though many skeptics said it couldn't be done'. The PPS also lists papers containing suspected plagiarized passages by automated synonymizing and automated paraphrasing. These papers feature 'tortured phrases' like 'fake neural organization' instead of 'artificial neural network'. The PPS uses a set of 276 tortured phrases as fingerprint-queries to identify (N=1694) problematic papers. The PPS queries Dimensions.ai regularly to identify new suspects among recently published/indexed papers. The PPS public online interface provides public users with the necessary information to assess suspected papers: access links (Dimension.ai record, DOI…), set of fingerprints found in the full text, link to existing PubPeer posts… User feedback is a means to both assess papers and identify new fingerprints/tortured phrases. Among the public list of suspected problematic papers (N=2088), 744 have been classified as problematic by various users and 1344 are still waiting for human assessments.

**Conclusion:**
The 'Problematic Paper Screener' automatically retrieves suspected published papers for scientists to reassess. The fingerprint-query approach is effective to identify computer-generated papers and synonymized plagiarism. In the future the approach will be tested for other problematic practices, such as the reporting of misidentified biological materials.

___________________
Guillaume Cabanac
University of Toulouse, Computer Science Department, IRIT UMR 5505 CNRS, 31062 Toulouse, France
E-mail: guillaume.cabanac@univ-tlse3.fr
ORCID: 0000-0003-3060-6241

Cyril Labbé
Univ. Grenoble Alpes, CNRS, Grenoble INP, LIG, 38000 Grenoble, France
E-mail: cyril.labbe@univ-grenoble-alpes.fr
ORCID: 0000-0003-4855-7038

Alexander Magazinov
Yandex, 82 Sadovnicheskaya str., Moscow 115035, Russia
E-mail: magazinov-al@yandex.ru
ORCID: 0000-0002-9406-013X